\documentclass[%
 reprint,
superscriptaddress,
 amsmath,amssymb,
aps,
prb,
]{revtex4-2}

\usepackage{graphicx}
\usepackage{dcolumn}
\usepackage{bm}
\usepackage{hyperref}

\begin{document}


\title{High-mobility two-dimensional electron gas in $\gamma$-Al$_2$O$_3$/SrTiO$_3$ heterostructures}

\author{Xiang-Hong Chen}
\affiliation{Tianjin Key Laboratory of Low Dimensional Materials Physics and Preparing Technology, Department of Physics, Tianjin University, Tianjin 300354, China}
\author{Zhi-Xin Hu}
\affiliation{Center for Joint Quantum Studies and Department of Physics, Tianjin University, Tianjin 300354, China}
\author{Kuang-Hong Gao}
\email[Corresponding author, e-mail: ]{khgao@tju.edu.cn}
\affiliation{Tianjin Key Laboratory of Low Dimensional Materials Physics and Preparing Technology, Department of Physics, Tianjin University, Tianjin 300354, China}
\author{Zhi-Qing Li}
\email[Corresponding author, e-mail: ]{zhiqingli@tju.edu.cn}
\affiliation{Tianjin Key Laboratory of Low Dimensional Materials Physics and Preparing Technology, Department of Physics, Tianjin University, Tianjin 300354, China}

\date{\today}

\begin{abstract}
The origin of the two-dimensional electron gas (2DEG) in the interface between $\gamma$-Al$_2$O$_3$ (GAO) and SrTiO$_3$ (STO) (GAO/STO) as well as the reason for the high mobility of the 2DEG is still in debate. In this paper, the electronic structures of [001]-oriented  GAO/STO heterostructures with and without oxygen vacancies are investigated by
first-principle calculations based on the density functional theory. The calculation results show that the necessary condition for the formation of 2DEG is that the GAO/STO heterostructure has the interface composed of Al and TiO$_2$ layers.
For the heterostructure without oxygen vacancy on the GAO side, the 2DEG originates from the polar discontinuity near the interface, and there is a critical thickness for the GAO film, below which the 2DEG would not present and the heterostructure exhibits insulator characteristics. For the case that only the GAO film contains oxygen vacancies, the polar discontinuity near the interface disappears, but the 2DEG still exists. In this situation, the critical thickness of the GAO film for 2DEG formation does not exist either. When the GAO film and STO substrate both contain oxygen vacancies, it is found that the 2DEG retains as long as the oxygen vacancies on the STO side are not very close to the interface. The low-temperature mobilities of the 2DEGs in these GAO/STO heterostructures are considered to be governed by the ionized impurity scattering, and $\sim$3 to $\sim$11 times as large as that in LaAlO$_3$/SrTiO$_3$ heterojunction. The high mobility of the 2DEG is mainly due to the small electron effective mass in GAO/STO heterostructure.
\end{abstract}

\maketitle


\section{INTRODUCTION}\label{SecI}

Since the report of existence of two-dimensional electron gas (2DEG) at the LaAlO$_3$/SrTiO$_3$ (LAO/STO) interfaces~\cite{ohtomo_high-mobility_2004}, great attention has been paid on the interfaces of STO-based heterostructures~\cite{nakagawa_why_2006,PhysRevLett.98.216803,kalabukhov_effect_2007,PhysRevLett.103.226802,PhysRevLett.105.236802,PhysRevLett.104.126803,gunkel_stoichiometry_2013,perna_conducting_2010,PhysRevB.86.085450,chen_high-mobility_2013,christensen_electric_2016,lee_origin_2013,PhysRevB.91.165118,PhysRevMaterials.4.016001,wolff_anisotropic_2017}. The 2DEGs confined in the STO-based heterostructures could reveal a range of unique phenomena including the existence of superconductivity at
low temperatures~\cite{reyren_anisotropy_2009,reyren_superconducting_2007}, electric-field-tuned metal–insulator
and superconductor–insulator phase transitions~\cite{caviglia_electric_2008,cen_nanoscale_2008,bell_dominant_2009,thiel_tunable_2006}, and coexistence of
ferromagnetism and superconductivity at the interfaces~\cite{bert_direct_2011,li_coexistence_2011}. These phenomena make STO-based heterostructures not only have potential applications in novel electronic devices but also be good model systems for fundamental research. Recently, a 2DEG with a low-temperature mobility of $\sim$$1.4\times10^5$\,cm$^2$/Vs, being much higher than that in LAO/STO interfaces, has been realized at the interface between STO and a heteroepitaxial spinel $\gamma$-Al$_2$O$_3$ (GAO) thin film~\cite{chen_high-mobility_2013}. This finding shows promise for the applications of the STO-based heterostructures in future electronic devices. Thus it is crucial to understand the formation mechanisms and the origin of the high mobility of the 2DEG at GAO/STO interfaces. However, there has been no consensus for the formation mechanisms of 2DEG in this heterostructure up to now. Some groups believe that the GAO in the heterostructures has polarity, and the charge accumulation is caused by the polar discontinuity at the interface~\cite{PhysRevMaterials.4.016001,PhysRevB.91.165118,wolff_anisotropic_2017}; while some groups argue that the formation of 2DEG is caused by oxygen vacancies at the interface~\cite{kormondy_quasi-two-dimensional_2015,schutz_microscopic_2017,lee_creation_2012,ngo_quasi-two-dimensional_2015}. Besides, the origin of the high mobility remains poorly understood as well~\cite{schutz_microscopic_2017,chen_room_2014}. To explore why high-mobility 2DEG can be formed at the GAO/STO interfaces, we systematically investigated the electronic structure of GAO/STO heterostructures via first-principles calculations. It is found that the polar discontinuity is the main mechanism for the formation of 2DEG in GAO/STO heterojunction when there is no oxygen vacancy on the GAO side; while the oxygen vacancies are the main source of the 2DEG for the heterojunction that contains oxygen vacancies on the GAO side. The reasons for the high mobility of the 2DEG in the GAO/STO heterojunctions are also disclosed.

\section{CALCULATION METHODS AND HETEROJUNCTION CONSTRUCTION}\label{SecII}
All calculations are carried out in the frame work of density-functional theory using a first-principle calculation software Vienna \emph{ab initio} simulation package. The generalized gradient approximation (GGA) parametrized by Perdew-Burke-Ernzerhof (PBE) plus the on-site coulomb interaction approach (GGA+$U$) was used for the exchange-correlation functional. For all heterostructures, a Hubbard interaction energy $U=8.5$\,eV for Ti 3$d$ orbitals was found to be appropriate~\cite{lee_charge_2008}. A cutoff energy of 500\,eV for plane wave basis set and a Monkhorst-Pack k-point mesh of $4\times4\times1$ were employed. The electronic energy convergence was set to be 10$^{-6}$\,eV, and a Gaussian smearing of 0.05\,eV was employed for density of states (DOS) calculations.

\begin{figure}[htp]
\includegraphics[scale=1]{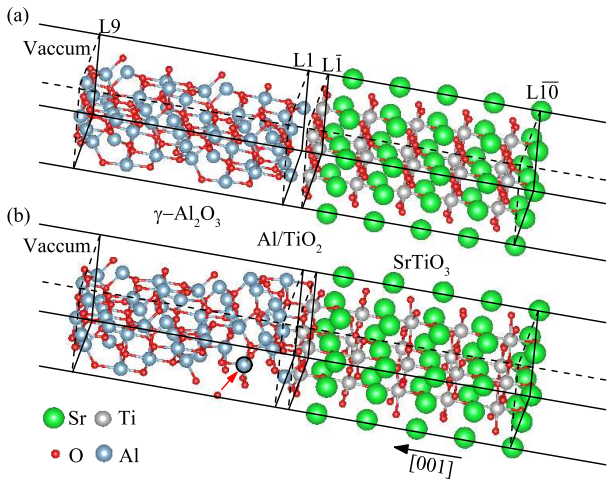}
\caption{Schematic structures of (GAO)$_{2.25}$/(STO)$_5$ heterostructures with Al and TiO$_2$ layers terminations (a) before, and (b) after atomic optimization. The atomic layers on the GAO side are labelled as L1, L2, $\cdots$, L9 (the adjacent Al and AlO atomic layers share the same number), and the atomic layers on the STO side are numbered as L$\bar{1}$, L$\bar{2}$, $\cdots$, L$\bar{10}$. For clarity, only L1, L9, L$\bar{1}$, and L$\bar{10}$ are shown in the diagram.}\label{geome-struc-GAO-STO}
\end{figure}

For GAO/STO heterostructures, considering the experimental results that the 2DEG can only be present in the interface of GAO and the TiO$_2$ terminated (001) STO substrate~\cite{wolff_anisotropic_2017,PhysRevB.99.134423}, we construct a $2\times 2$ in-plane STO supercell with TiO$_2$ layer terminated on the GAO side and SrO layer terminated on the vacuum side to simulate the [001]-oriented STO substrate. Along the [001] direction, the supercell contains 5 TiO$_2$ and 5 SrO layers, respectively (see Fig.~\ref{geome-struc-GAO-STO}). Usually, GAO belongs to the defective spinel structure with the space group $Fd\bar{3}m$. In an ideal spinel structure, a unit cell (uc) contains 24 Al atoms and 32 oxygen atoms (8 Al atoms in the tetrahedral sites, 16 Al atoms in the octahedral sites, and 32 O atoms forming a fcc array), and the ratio of cation to anion is $3:4$. In order to obtain stoichiometric Al$_2$O$_3$, 8/3 Al vacancies need to be created per spinel unit cell in average. Therefore, at least 0.75 spinel unit cells are required for GAO (001) surface to match the Al$_2$O$_3$ stoichiometric and vacancy distribution requirements (two Al vacancies). In our calculations, the thickness of GAO slabs varies from 0.75 to 3\,uc. For the distribution of Al vacancies in GAO slab, three aspects are considered~\cite{gutierrez_theoretical_2001,vijay_structure_2002,PhysRevB.70.125402,PhysRevB.79.235410,menendez-proupin_electronic_2005}: (1) Al vacancies preferentially occupy the octahedral positions; (2) the distance between Al vacancies is as far as possible to ensure that the vacancies are distributed as homogeneously as possible in the lattice; and (3) Al vacancies are distributed as deep as possible inside GAO rather than on its surface.

Since the structure of GAO can be viewed as an alternating stacking sequence of Al atomic layer and AlO atomic layer along the [001] direction, both Al/TiO$_2$ and AlO/TiO$_2$ interfaces were considered,  respectively. Finally, a series of configurations of (GAO)$_n$/(STO)$_5$ ($n$=0.75, 1.5, 2.25, and 3) with Al/TiO$_2$ or AlO/TiO$_2$ interfaces were constructed (i.e., there are eight configurations). A vacuum layer with thickness of 15\,${\rm {\AA}}$ is inserted between neighbor slabs. Structural optimization was done for all the configurations. All atoms, including the surface layer, are optimized until the residual stress is less than 0.03\,eV/${\rm {\AA}}$. It is found that the configurations with AlO/TiO$_2$ interface were not significantly changed before and after optimization. However, the case of the configurations with Al/TiO$_2$ interface is different. Figure~\ref{geome-struc-GAO-STO}(a) and \ref{geome-struc-GAO-STO}(b) show the interface geometry of (GAO)$_{2.25}$/(STO)$_5$ heterostructure with Al/TiO$_2$ interface before and after optimization. For simplicity, we define (GAO)$_{2.25}$/(STO)$_5$ heterostructure with Al/TiO$_2$ interface as Configuration 1 (Conf.~1). We find that the structure of GAO near the interface in the optimized Conf.~1 is distorted compared with the structure before optimization. As a consequence, the positions of atoms in the three Al layers and AlO layers near the interface have been changed. For example, the Al atom (pointed out by arrow) in the L2 AlO layer should be at the center of the octahedron, while it deviates from that position after optimization. Although the structure distortion occurs in the GAO slab near the interface, the alternating structure of octahedral cation and tetrahedral cation along the [001] direction is still retained, which is consistent with the experimental results~\cite{kormondy_quasi-two-dimensional_2015,lu_spectrum_2016}.

\begin{figure}
	\includegraphics[scale=0.75]{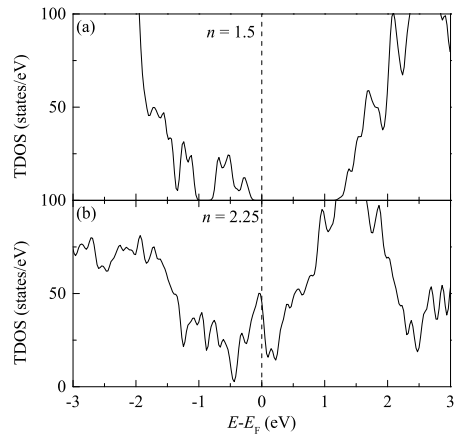}
	\caption{The DOS of (GAO)$_n$/(STO)$_5$ heterostructures with Al/TiO$_2$ interface for (a) $n=1.5$, and (b) $n=2.25$. }\label{Fig-DOS-Total}
\end{figure}

\begin{figure*}[htp]
	\includegraphics[scale=0.75]{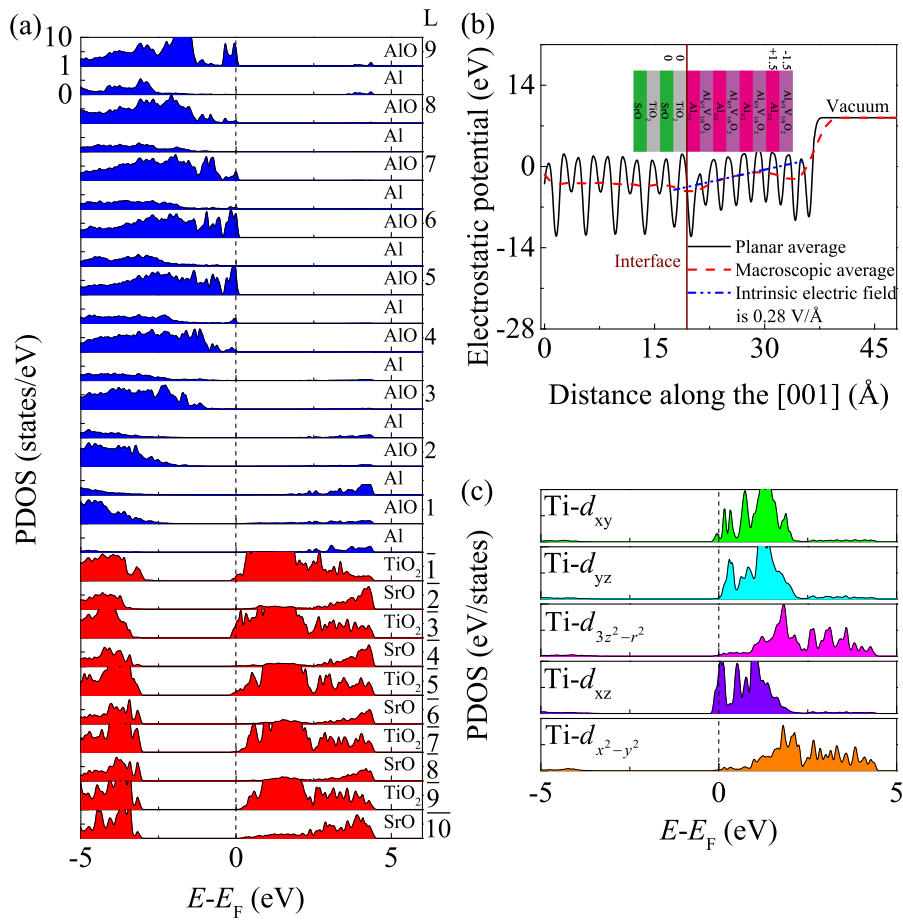}
	\caption{(a) The partial density of states projected onto atomic planes in the STO and GAO of Conf.~1. (b) The plane average electrostatic potential (solid line) and macroscopic average electrostatic potential (dashed line) of Conf.~1. Inset: a sketch of the GAO/STO heterostructure. (c) The orbital resolved partial DOS for L$\bar{3}$ TiO$_2$ layer in STO.}\label{Conf1-PDOS-POT}
\end{figure*}

\section{RESULTS AND DISCUSSIONS}\label{SecIII}
\subsection{GAO/STO heterostructures without oxygen vacancies}
Firstly, we calculated the electron DOS of the GAO/STO heterostructures with the eight configurations mentioned above. It is found that the Fermi levels of the heterostructures with AlO/TiO$_2$ interface lie in the bandgap no matter how thick the GAO layer is. Thus the AlO-layer-terminated heterostructures are always in the insulating state. The situation on GAO/STO heterostructures with Al/TiO$_2$ interface is different. Figure~\ref{Fig-DOS-Total}(a) and \ref{Fig-DOS-Total}(b) show the total DOS of the (GAO)$_n$/(STO)$_5$ heterostructures with Al/TiO$_2$ interface for $n=1.5$ and 2.25, respectively. The Fermi level for the $n=1.5$ heterostructure is in the bandgap, indicating that the $n=1.5$ heterostructure is an insulator in the band structure. The band structure of the $n=0.75$ heterostructure is similar to that of the $n=1.5$. In contrast, the gap vanishes entirely and apparent nonzero DOS can be observed near the Fermi level for the $n=2.25$ heterostructure [Fig.~\ref{Fig-DOS-Total}(b)]. That is, the $n=2.25$ heterostructure possesses band structure of metals and similar phenomenon is also observed in the $n=3$ heterostructure. Therefore, the 2DEG will be formed in GAO/STO heterostructure with Al/TiO$_2$ interface when the thickness of GAO is greater than 1.5\,uc ($n>1.5$), and the critical thickness of GAO for the presence of 2DEG in GAO/STO heterostructure is $\sim$1.5\,uc. Since the results in the $n=3$ heterostructure are similar to those in the $n=2.25$, we will systematically present the results and discussions concerning the GAO/STO heterostructures with Al/TiO$_2$ interface and $n=2.25$ below.

Figure~\ref{Conf1-PDOS-POT}(a) shows the partial density of states projected onto atomic planes in the STO and GAO of the $n=2.25$ heterostructure (Conf.~1). From this figure, one can see that the states at the top of the valence band are mainly contributed by the AlO layers, and the related valence-band edge gradually shifts upward with increasing layer number from L1 to L5 layers (i.e., the farther the AlO layer is from the interface, the greater the related valence-band edge moves upward). On the other hand, the conduction-band edge of L$\bar{1}$ to L$\bar{3}$ TiO$_2$ layers is lower than the Fermi level, thus the valence band of L5 to L9 AlO layers has an overlap with the conduction band of L$\bar{1}$ to L$\bar{3}$ TiO$_2$ layers. These band overlaps cause the electrons to be transferred from the former to the latter, resulting in incompletely filled bands, and hence the 2DEG near the interface.

As for the origin of the band overlap, the polar discontinuity at GAO/STO interfaces is the most likely candidate. We have known that the GAO slab in optimized Conf.~1 maintains the alternating structure of octahedral cation and tetrahedral cation along the [001] direction [Fig.~\ref{geome-struc-GAO-STO}(b)]. The homogeneous distribution of aluminum vacancies on the octahedral site leads to each AlO layer having a $-1.5e$ charge, and then each tetrahedral layer containing only aluminum cations has a charge of $+1.5e$. Therefore, the GAO can be interpreted as stacking of atomic layers with an alternating charge of $\sigma=\pm1.5e$~\cite{CHRISTENSEN2017887} [see inset of Fig.~\ref{Conf1-PDOS-POT}(b)], and the GAO is a polar crystal with nonvanishing electric dipole moment in the [001] direction~\cite{CHRISTENSEN2017887,tasker_stability_1979}. The main panel of Fig.~\ref{Conf1-PDOS-POT}(b) shows the plane average (solid curve) and macroscopic average (dashed curve) electrostatic potential of Conf.~1 as a function of the position in the direction normal to the interface. The potential in GAO increases gradually along $z$ direction due to the alternating polarity of the atomic layers, indicating the presence of a polar electric field in GAO. Consequently, the polar electric field causes an upward shift of the valence-band edge. By a linear fitting the macroscopic average electrostatic potential~\cite{harrison_polar_1978}, a polar electric field of 0.28\,V/${\rm {\AA}}$ is obtained. Using the electric field of 0.28\,V/${\rm {\AA}}$ and the GGA+U band gap of 3.0\,eV for STO, one can obtain the critical thickness of GAO that is $\sim$1.4\,uc. This value is close to our result mentioned above. If the electric field of 0.28\,V/${\rm {\AA}}$ was completely compensated, 0.92 electrons per unit-cell area (e/A) (equivalently, carrier concentration value is $1.47\times10^{14}\,{\rm {cm}}^{-2}$) should be transferred from GAO to STO. Integrating the DOS spectra from the bottom of the conduction band to the Fermi level, one can  obtain the carrier concentration $n$ of the heterojunction. The carrier concentration in the slab with Conf.~1 is $1.14\times10^{14}\,{\rm {cm}}^{-2}$, which is close to the density of the compensating electrons mentioned above. This result confirms that the polar discontinuity induces the formation of 2DEG in Conf.~1, which is similar to the case in LAO/STO heterostructures~\cite{behtash_polarization_2016,son_density_2009}. The slight deviation between the two values could be caused by the ionic polarization that partially screens the polar electric field in GAO~\cite{li_formation_2011}. And we find that 66.4\% of the electrons occupy the Ti-3$d$ orbitals of the L$\bar{3}$ TiO$_2$ layer. Therefore, the metallic state of Conf.~1 is mainly contributed by the L$\bar{3}$ TiO$_2$ layer.

\begin{figure}
\includegraphics[scale=1]{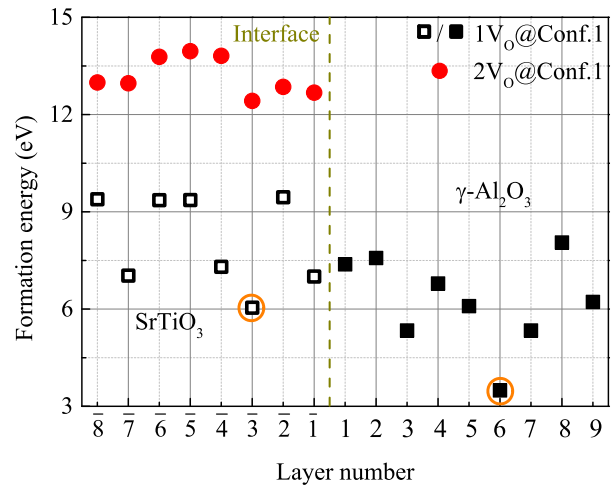}
\caption{Formation energies of oxygen vacancies at different atomic layers in Conf.~1. The legend of 1V$_{\rm O}$@Conf.~1 represents one oxygen vacancy in the Conf.~1, the other follows the same convention. For the Conf.~1 containing two oxygen vacancies, one is fixed in the L6 AlO layer and the other varies at different atomic layers on the STO side. For the Conf.~1 containing one oxygen vacancy, the circles mark the positions with the lower formation energy of the oxygen vacancy on the STO side and the GAO side.}\label{VO-in-Conf1-Ef}
\end{figure}

Figure~\ref{Conf1-PDOS-POT}(c) shows the orbital resolved partial DOS for the L$\bar{3}$ TiO$_2$ layer. The electrons transferred from the AlO layer mainly occupy the Ti-3$d_{xz}$ orbitals (about 77.1\% of the electrons occupy these orbitals) and partially occupy Ti-3$d_{xy}$ and Ti-3$d_{yz}$ orbitals. This is different from LAO/STO heterostructures, where the preferential occupation of 2DEG is the in-plane Ti-3$d_{xy}$ orbitals~\cite{chikina_orbital_2018}. This anomalous behavior could be attributed to the contact of the interfacial Ti cations with the tetrahedral Al cations. The Al/TiO$_2$ interface breaks the perovskite lattice symmetry and changes the crystal field around the Ti ions near the interface. As a result, the energy of $d_{xz}$/$d_{yz}$ orbitals is reduced and these orbitals become the preferable states for the electrons at the spinel-perovskite interface. The experimental results of Chikina~\emph{et al.}~\cite{chikina_band-order_2021} and Cao~\emph{et al.}~\cite{cao_anomalous_2016} also indicate that the energy of the $d_{xz}/d_{yz}$  orbitals is less than that of the $d_{xy}$ orbitals in GAO/STO heterostructures.

\begin{figure*}
	\centering 	
	\includegraphics[scale=0.75]{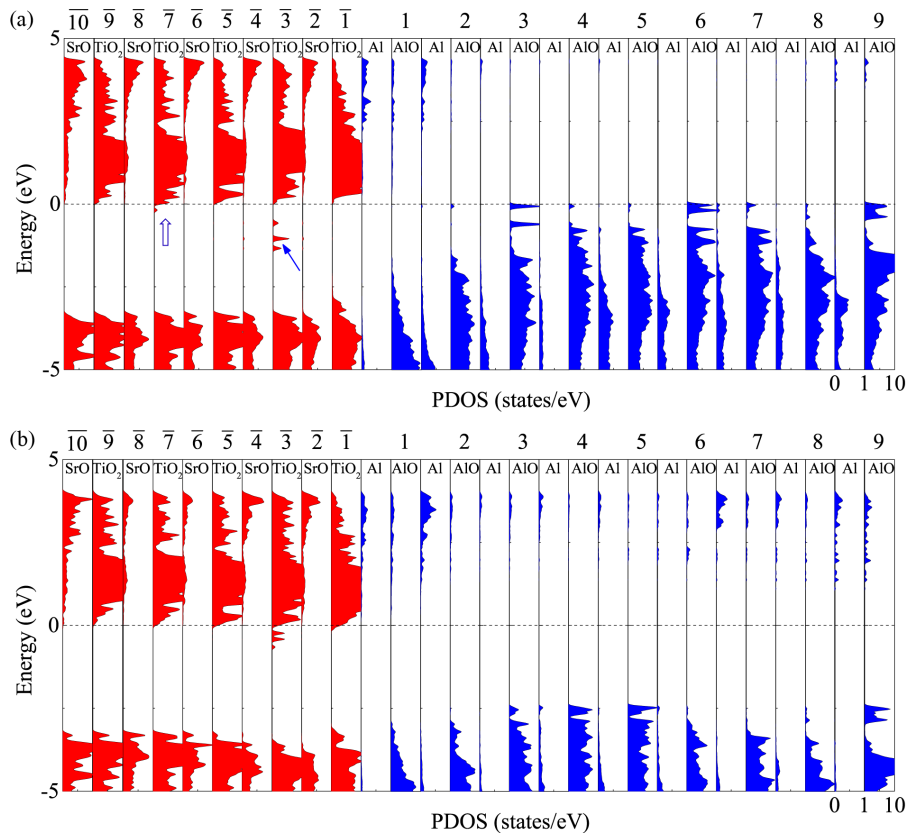}
	\caption{The partial density of states projected onto atomic planes in the STO and GAO for (a) the ($\gamma$-Al$_2$O$_3$)$_{2.25}$/(SrTiO$_{2.95}$)$_5$ heterostructures with one oxygen vacancy on L$\bar{3}$ TiO$_2$ layer (Conf.~2), and (b) the ($\gamma$-Al$_2$O$_{2.96}$)$_{2.25}$/(SrTiO$_{3}$)$_5$ heterostructures with one oxygen vacancy on L6 AlO layer (Conf.~3). The hollow arrow in (a) marks the electron accumulation layer and the solid arrow points out the energy states of oxygen vacancies.}\label{VO-in-3TiO2/6AlO-PDOS-POT}
\end{figure*}

\subsection{Influence of oxygen vacancies on the electronic structure of GAO/STO heterostructures}
As mentioned in the Sec.~\ref{SecI}, some researchers have suggested that the formation of 2DEG in GAO/STO heterostructures is caused by oxygen vacancies. Therefore, the effect of oxygen vacancies on the formation of 2DEG in GAO/STO heterostructures is further investigated. Considering that oxygen vacancies could be located in AlO, SrO, and TiO$_2$ layers, we construct ($\gamma$-Al$_2$O$_{2.96}$)$_{2.25}$/(SrTiO$_3$)$_5$ heterostructures and ($\gamma$-Al$_2$O$_3$)$_{2.25}$/(SrTiO$_{2.95}$)$_5$ heterostructures (each configuration contains only one oxygen vacancy in the supercell). Firstly, we obtained the optimized structures of these configurations. All crystal structures experience similar structural distortions as that in Conf.~1. Considering the GAO with Al layer termination at the interface, we calculate the formation energy ($E_{f}$) of oxygen vacancies using the relation,
\begin{equation}\label{VO-Ef}
	E_{f}=E_{\rm {V}}-(E_{0}-n_V\mu_{\rm {O}}),
\end{equation}
where $E_{\rm {V}}$ and $E_0$ are the total energies of the interface structure with and without oxygen vacancies, $n_V$ is the numbers of oxygen vacancies, and $\mu_{\rm {O}}$ is the chemical potential of oxygen atom. Figure~\ref{VO-in-Conf1-Ef} shows the formation energy as a function of the position of the oxygen vacancy. The square symbol represents the formation energy of the oxygen vacancy for the supercell containing only one oxygen vacancy. Inspection of Fig.~\ref{VO-in-Conf1-Ef} indicates that there are two energy minima, one lies in the L$\bar{3}$ TiO$_{2}$ layer and the other in the L6 AlO layer. The partial density of states projected onto atomic planes in the STO and GAO for oxygen vacancy located in the L$\bar{3}$ TiO$_{2}$ layer (designated as Conf.~2) and the L6 AlO layer (designated as Conf.~3) are plotted in Fig.~\ref{VO-in-3TiO2/6AlO-PDOS-POT} (a) and \ref{VO-in-3TiO2/6AlO-PDOS-POT} (b), respectively. For Conf.~2, the top of the valence band also overlaps with the bottom of the conduction band, which is similar to that in Conf.~1. Thus the 2DEG would also appear in the heterostructure with structure of Conf.~2. However, the 2DEG in Conf.~2 is mainly located in the L$\bar{7}$ TiO$_2$ layer, which is different from that in Conf.~1. In addition, the electrons transferred from the AlO layer mainly occupy the Ti-3$d_{xy}$ orbitals. This is expected because the L$\bar{7}$ TiO$_2$ layer is far from the Al/TiO$_2$ interface and located deeply in the STO substrate. The carrier concentration of Conf.~2 is $5.18\times10^{13}\,{\rm {cm}}^{-2}$, which is less than the concentration to fully compensate the polar electric field of GAO ($1.47\times10^{14}\,{\rm {cm}}^{-2}$). Two factors could be responsible for this phenomenon: (1) the oxygen vacancies in the L$\bar{3}$ layer could affect the lattice distortion near the interface and thus reduce the polar electric field; (The calculated polar field of Conf.~2 is $\sim$0.24\,V/${\rm {\AA}}$, which is less than that of Conf.~1. Correspondingly the critical thickness of GAO film for the existence of 2DES should be $\sim$1.6\,uc.) (2) the electrons released by the oxygen vacancies are found to form band gap states [marked by the solid arrow in Fig.~\ref{VO-in-3TiO2/6AlO-PDOS-POT}(a)] and consequently have no contribution to the carrier concentration.

\begin{table}
	\caption{\label{table-I}Calculated carrier concentration $n$, average electron effective mass $\overline{m^*}/m_e$, $k\alpha$, and electron mobility $\mu$ of Conf.~1, 2, 3, 5, and (LAO)$_5$/(STO)$_5$  heterostructures.}
	\begin{ruledtabular}
		\begin{tabular}{cccccccc}
			Configuration & $n$\,(cm$^{-2}$) & $\overline{m^*}/m_{\rm {e}}$ & $k\alpha$ & $\mu$\,(cm$^2$/Vs) \\ \hline
			Conf.~1 & $1.14\times10^{14}$ & 1.20 & 12.27 &$5.30\times10^4$  \\
			Conf.~2 & $5.18\times10^{13}$ & 0.69 & 14.22 &$1.52\times10^5$  \\
			Conf.~3 & $2.05\times10^{14}$ & 0.62 & 18.87 &$1.71\times10^5$  \\
			Conf.~5 & $2.60\times10^{14}$ & 0.60 & 19.94 &$1.80\times10^5$  \\
			(LAO)$_5$/(STO)$_5$ & $1.32\times10^{12}$ & 2.81 & 3.630 &$1.74\times10^4$ \\
		\end{tabular}
	\end{ruledtabular}
\end{table}

\begin{figure}[b]
\includegraphics[scale=1]{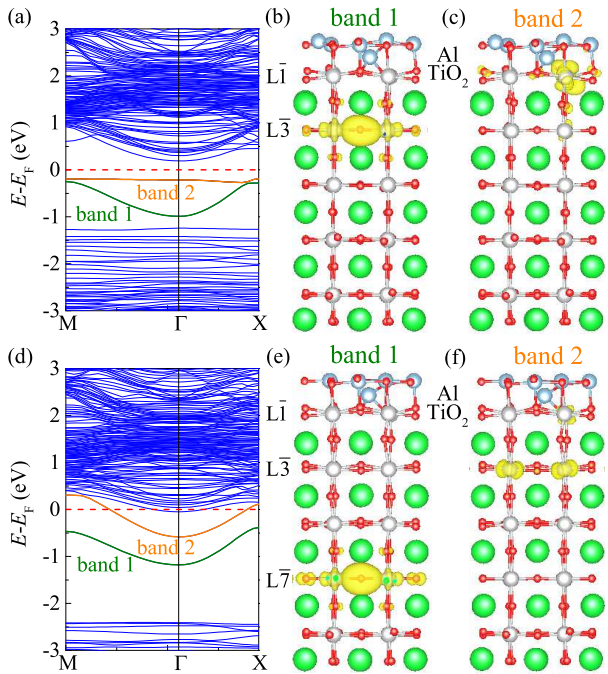}
\caption{(a) The band structure of ($\gamma$-Al$_2$O$_{2.96}$)$_{2.25}$/
(SrTiO$_{2.95}$)$_5$ heterostructure with one oxygen vacancy in the L6 AlO layer and the other in the L$\bar{3}$ TiO$_2$ layer. (b) and (c) partial charge density isosurfaces for the band~1 and band~2. (d) The band structure of ($\gamma$-Al$_2$O$_{2.96}$)$_{2.25}$/(SrTiO$_{2.95}$)$_5$ heterostructure with one oxygen vacancy in the L6 AlO layer and the other in the L$\bar{7}$ TiO$_2$ layer. (e) and (f) partial charge density isosurfaces for the band~1 and band~2. The isosurface value is 0.045\,e/${\rm {\AA}^3}$. }\label{VO-in-6AlO-3/7TiO2-band-charge density}
\end{figure}

For Conf.~3, there is no any overlap between the top of the valence band and the bottom of the conduction band [see Fig.~\ref{VO-in-3TiO2/6AlO-PDOS-POT}(b)]. However, the Fermi level shifts upward into the conduction band, and the slab still reveals metallic state in band structure. Therefore, the 2DEG in Conf.~3 does not originate from polar discontinuity at the GAO/STO interface but from the oxygen vacancies in the GAO (similar to modulation doping). An oxygen vacancy in the GAO/STO supercell could release one or two electrons to the STO~\cite{PhysRevLett.111.217601,PhysRevB.92.115112}. This concentration of the transferred electron should vary from $1.57\times10^{14}\,{\rm {cm}}^{-2}$ to $3.14\times10^{14}\,{\rm {cm}}^{-2}$. The carrier concentration of Conf.~3 is $2.05\times10^{14}\,{\rm {cm}}^{-2}$ (Table~{\ref{table-I}}), indicating a partial ionization of the oxygen vacancies. The electrons transferred from the oxygen vacancy in the GAO are mainly accumulated in the L$\bar{3}$ TiO$_2$ layer. The value of the carrier concentration in Conf.~3 is greater than the concentration to fully compensate the polar electric field of GAO ($1.47\times10^{4}\,{\rm {cm}}^{-2}$, as mentioned above). Thus, the polar electric field of GAO could be completely compensated. This suggests that the formation of 2DEG in GAO/STO heterostructures would be irrelevant to the thickness of the GAO slab. Then, the electronic structures of ($\gamma$-Al$_2$O$_{2.87}$)$_{0.75}$/(SrTiO$_3$)$_5$ and ($\gamma$-Al$_2$O$_{2.94}$)$_{1.5}$/(SrTiO$_3$)$_5$ (containing only one oxygen vacancy in the supercell on GAO side) are also calculated, and both of them exhibit metallic characteristics. Thus, there will be no critical thickness of GAO for the presence of 2DEG in the GAO/STO heterostructures when the oxygen vacancies are located on the GAO side. Experimentally, Chen~\emph{et al.} performed during-growth transport measurements for LAO/STO and related heterostructures and found that an amorphous film of less than 0.04\,nm monolayer coverage on STO can already generate a highly conducting 2DEG~\cite{chen_unravelling_2019}. The results demonstrate that the oxygen vacancies in GAO film grown on STO play an important role in generating 2DEG. We also calculate the electronic structures of the GAO/STO heterostructures with AlO and TiO$_2$ terminations and one oxygen vacancy in the supercell, and find that all configurations exhibit insulator characteristics in band structure.

Now, we consider the situation that in the supercell both the GAO and the STO sides contain one oxygen vacancy, respectively. As the formation energy of oxygen vacancy in the L6 AlO layer is the lowest (the filled square symbols in Fig.~\ref{VO-in-Conf1-Ef}), we fix one oxygen vacancy in the L6 AlO layer and change the position of the other in the STO side for the ($\gamma$-Al$_2$O$_{2.96}$)$_{2.25}$/(SrTiO$_{2.95}$)$_5$ heterostructures. The formation energy variation with the position of oxygen vacancy is also shown in Fig.~\ref{VO-in-Conf1-Ef} (the filled circle symbols), from which one can see that the formation energy does not change greatly with the change of oxygen vacancy position. This means that oxygen vacancies can be formed in any atomic layer on the STO side under low oxygen pressure.

Next, we focus on the electronic structures of ($\gamma$-Al$_2$O$_{2.96}$)$_{2.25}$/(SrTiO$_{2.95}$)$_5$ heterostructures. It is found that the ($\gamma$-Al$_2$O$_{2.96}$)$_{2.25}$/(SrTiO$_{2.95}$)$_5$ heterostructures show insulator characteristics in band structure when the oxygen vacancy is located in L$\bar{1}$, L$\bar{2}$, and L$\bar{3}$ layers, while for other cases, the heterostructures exhibit metallic characteristics. We take two configurations, in which the oxygen vacancy in the L$\bar{3}$ TiO$_2$ layer (designated as Conf.~4) and L$\bar{7}$ TiO$_2$ layer (designated as Conf.~5), respectively, as examples to present the detailed results. The band structure of Conf.~4 along high symmetry lines in the Brillouin zone is plotted in Fig.~\ref{VO-in-6AlO-3/7TiO2-band-charge density}(a). The Fermi level lies in the bandgap, indicating that the 2DEG cannot be present in this configuration. Below the Fermi level, there are two discrete energy bands, which are designated as band~1 and band~2 [as indicated in Fig.~\ref{VO-in-6AlO-3/7TiO2-band-charge density}(a)] for convenience. Figure~\ref{VO-in-6AlO-3/7TiO2-band-charge density}(b) shows the k-resolved charge density isosurfaces for band~1 in the entire two-dimensional Brillouin zone. The charge density of band~1 exhibits a typical Ti-3$d$ $e_g$-based bonding orbital characteristic, suggesting that the electrons are strongly localized at the oxygen vacancy and its adjacent Ti atoms [see L$\bar{3}$ layer in Fig.~\ref{VO-in-6AlO-3/7TiO2-band-charge density}(b)]. It is indicated that band~1 corresponds to the impurity band introduced by oxygen vacancies, while band~2 is introduced by electrons transferred from GAO to STO, and the flat band shows that the electron effective mass is large. Figure~\ref{VO-in-6AlO-3/7TiO2-band-charge density}(c) shows the  k-resolved charge density isosurfaces for band~2. The charge density is located on one of the Ti atoms in the L$\bar{1}$ TiO$_2$ layer, showing a rather complex $d$-orbital composition. It could be composed of $d_{xy}$, $d_{xz}$, $d_{yz}$, $d_{3z^2-r^2}$, and $d_{x^2-y^2}$ orbitals. Thus, the electrons transferred from GAO to STO form localized states. In contrast to Conf.~2 with only one oxygen vacancy in L$\bar{3}$ layer, no 2DEG is formed in Conf.~4 with an additional oxygen vacancy in the L6 layer. In fact, the additional oxygen vacancy could affect the crystal field near the Al/TiO$_2$ interface, which in turn causes the electrons transferred from GAO to STO to be localized.

Figure~\ref{VO-in-6AlO-3/7TiO2-band-charge density}(d) illustrates the band structure of Conf.~5. Unlike the electronic structure of Conf.~4, the Fermi level for Conf.~5 shifts upward into the conduction band, indicating that the 2DEG forms in heterostructures with this configuration. The bands introduced by the oxygen vacancies in STO and the electrons transferred from GAO to STO are also labeled as band~1 and band~2, respectively. Figure~\ref{VO-in-6AlO-3/7TiO2-band-charge density}(e) and \ref{VO-in-6AlO-3/7TiO2-band-charge density}(f) show the k-resolved charge density isosurfaces for the two bands. From Fig.~\ref{VO-in-6AlO-3/7TiO2-band-charge density}(e) one can see that the electrons in L$\bar{7}$ are localized at the oxygen vacancy and its adjacent Ti atoms, which is similar to that in L$\bar{3}$ layer of Conf.~4. Thus the oxygen vacancies in L$\bar{7}$ layer of STO are deep donors. Inspection of Fig.~\ref{VO-in-6AlO-3/7TiO2-band-charge density}(f) indicates that the electrons transferred from GAO to STO are mainly located in the L$\bar{3}$ TiO$_2$ layer, occupy the Ti-3$d_{xy}$ orbitals, and are uniformly distributed on each Ti atom. These electrons can only move freely in the L$\bar{3}$ TiO$_2$ layer, suggesting that the electrons transferred from GAO to STO are confined in the L$\bar{3}$ TiO$_2$ layer and form a 2DEG. It should be noted that a few electrons are localized on a Ti atom in the L$\bar{1}$ layer. The corresponding charge density isosurface region is far less than that of Conf.~4. Compared with the oxygen vacancies in the L$\bar{3}$ layer, the vacancies in the L$\bar{7}$ layer are farther away from the interface and have a weak effect on the crystal field near the Al/TiO$_2$ interface. Therefore, the electronic structure of Conf.~5 is similar to that of Conf.~3. The above results reveal that the formation of 2DEG in the GAO/STO heterostructures is closely related to the position of oxygen vacancies.

\subsection{The electron mobilities in GAO/STO heterostructures}
In the framework of free-electron-like model, the mobility can be expressed as $\mu=e\left \langle \tau \right \rangle /m^{*}$, where $e$, $\tau$, and $m^{*}$ are the elementary charge, relaxation time, and electron effective mass, respectively. Besides the relaxation time, the effective mass plays an important role in determining the electron mobility. From the energy versus wave vector dispersion near the bottom of the conduction band, we obtain the electron effective masses along the $\Gamma$-M direction and the $\Gamma$-X direction. Then the average effective masses can be obtained using $\overline{m^*}=\sqrt{m_{\Gamma-{\rm {M}}}^{*}m_{\Gamma-{\rm {X}}}^{*}}$. The average effective masses for Confs.~1, 2, 3, and 5 are listed in Table~{\ref{table-I}}. For comparison, we also calculated the band structure of n-type LAO/STO heterostructure with configuration (LAO)$_5$/(STO)$_5$ (i.e., the five LAO layers are stacked on the five STO layers). Our results concerning the energy band of (LAO)$_5$/(STO)$_5$ are almost identical to those in Ref.~\cite{son_density_2009}. The lowest band near the $\Gamma$ point is composed of $d_{xy}$ orbitals, and $d_{yz}$ subbands show small dispersions from $\Gamma$ to X points. The average effective mass of (LAO)$_5$/(STO)$_5$ is also given in Table~{\ref{table-I}}. For the GAO/STO heterostructures, the effective mass of Conf.~1 is the largest, which is about 3/7 of that in (LAO)$_5$/(STO)$_5$. While there is little difference in effective masses of Confs.~2, 3, and 5, which is about one-forth of that in (LAO)$_5$/(STO)$_5$. Hence, the smaller effective masses must be one of the important reasons for the high mobility of 2DEG in GAO/STO heterostructures.

According to Christensen~\emph{et al.}~\cite{christensen_electron_2018}, the electron mobility in GAO/STO heterostructures is governed by ionized impurity scattering at low temperature regime ($T<5$\,K). The ionized impurity scattering has been treated theoretically by Mansfield~\cite{tufte_electron_1967,chattopadhyay_electron_1981,frederikse_hall_1967,verma_intrinsic_2014}. When the screen radius ($\alpha$) is far greater than the wavelength of electrons (i.e., $k\alpha\gg1$ with $k$ being the wave vector), the ionized-impurities-scattering limited electron mobility can be expressed as~\cite{tufte_electron_1967,chattopadhyay_electron_1981,frederikse_hall_1967,verma_intrinsic_2014}
\begin{equation}\label{mobility-low-temp}
	\mu =\frac{3\epsilon^2h^3n}{16\pi^2e^3Z^2N_im^{*2}g(b)}
\end{equation}
with
\begin{equation}\label{gb}
	g(b)=\ln(1+b)-\frac{b}{1-b}
\end{equation}
and
\begin{equation}\label{b}
	b=\frac{h^2\epsilon}{2e^2m^*} \left(\frac{3n}{8\pi}\right)^{1/3},
\end{equation}
where $\epsilon$ is the dielectric constant, $h$ is Planck's constant, $n$ is the carrier concentration, $Z$ is the charge of donors and $N_i$ is the concentration of charged donor centers. The screening radius $\alpha$ can be calculated by the Thomas-Fermi expression~\cite{verma_intrinsic_2014,frederikse_hall_1967,march_thomas-fermi_1957},
\begin{equation}\label{Thomas-Fermi}
	\alpha = \left[\frac{\hbar^2\epsilon}{4m^*e^2}\left(\frac{\pi}{3n}\right)^{1/3}\right]^{1/2},
\end{equation}
where $\hbar$ is the Planck's constant divided by 2$\pi$. The values of $\alpha$ were obtained by substituting the $n$, $\overline{m^*}$, and $\epsilon=18000$ (STO) into Eq.~(\ref{Thomas-Fermi}). Taking $k=k_{\rm {F}}$ with $k_{\rm {F}}$ being the Fermi wave vector, we obtain the values of $k\alpha$ and list them in Table~{\ref{table-I}}. Checking Table~{\ref{table-I}} indicates that the condition $k\alpha\gg1$ is satisfied for each listed configuration. Since the charge donors are partially ionized, the values of $Z$ and ${N_i}$ are taken as 1 and $n$, respectively. Thus the electron mobility for each configuration is obtained and listed in Table~{\ref{table-I}}. The electron mobility in Conf.~1 is about 3 times as large as that in (LAO)$_5$/(STO)$_5$, while the electron mobilities in heterostructures with Confs.2, 3, and 5 are $\sim$9, 10, and 11 times as large as that in (LAO)$_5$/(STO)$_5$. In addition, for the GAO/STO heterostructures, it is worth noting that there is a certain distance between the electrons (forming 2DEG) and their donors in all configurations with 2DEGs. For example, in Conf.~3, the 2DEG lies in the L$\bar{3}$ TiO$_2$ layer, which is 1.6\,nm away from the oxygen vacancy lying in the L6 AlO layer. This is equivalent to realizing the modulation doping~\cite{schutz_microscopic_2017}, which is a general way to suppress the donor scattering and improves the mobility in semiconductor technology. Therefore, the electron mobility in each GAO/STO heterostructure should be greater than the calculated value in Table~{\ref{table-I}}. The modulation doping, together with the smaller effective masses, explains why the higher mobility 2DEGs exist in the GAO/STO heterostructures (compared to LAO/STO heterostructures).

\section{CONCLUSIONS}\label{SecIV}
In summary, the electronic structures of the [0\,0\,1]-oriented GAO/STO heterostructures with TiO$_2$ layer terminated on the interfaces are systematically investigated by the first-principle calculations. It is found that the necessary condition for the existence of 2DEG in the heterostructure is the device possessing interface with TiO$_2$ and Al layers terminations. For GAO/STO heterojunction without oxygen vacancy on the GAO side, the polar discontinuity near the interface is the main mechanism for the occurrence of 2DEG. When the oxygen vacancies are introduced into the GAO films, polar electric field near the interface disappears and the oxygen vacancies become the main source of the 2DEG.
In GAO/STO heterostructure the ionized impurity scattering is considered to govern the low-temperature mobility of the 2DEG, and it is found that the mobility in GAO/STO heterostructure with proper configurations can be one order of magnitude higher than that in LAO/STO heterostructure. The high mobility of the 2DEG is mainly due to the small electron effective mass in GAO/STO heterostructure.
\begin{acknowledgments}
The calculation was conducted on the CJQS-HPC platform at Tianjin University. This work is supported by the National Natural Science Foundation of China through Grant Nos. 12174282 and 11774253.
\end{acknowledgments}

\bibliography{reference}

\end{document}